\documentclass[10pt,twocolumn]{IEEEtran}

\usepackage{graphicx}
\usepackage{times}
\usepackage{color}
\usepackage{fix2col}
\usepackage{amsmath,amsfonts,amssymb}

\newcommand{\eq}{\triangleq}

\newcommand{\field}[1]{\mathbb{#1}}
\newcommand{\R}{\field{R}}

\newcommand{\U}{{\mathcal{U}}}

\newcommand{\diag}{\mathrm{diag}}
\newcommand{\blockdiag}{\mathrm{blockdiag}}
\newcommand{\vc}[1]{{\boldsymbol{#1}}}

\newtheorem{thm}{Theorem}

\newtheorem{rem}[thm]{Remark}
\newtheorem{problem}[thm]{Problem}

\title{\LARGE \bf
Hands-Off Control as Green Control}

\author{Masaaki Nagahara, Daniel E. Quevedo, Dragan Ne\v{s}i\'{c}%
\thanks{
This research is supported in part by the JSPS Grant-in-Aid for Scientific Research (C) No.~24560543,
and also
Australian Research Council's
Discovery Projects funding scheme (project number DP0988601). 
}%
\thanks{
M. Nagahara is with
 Graduate School of Informatics, Kyoto
 University, Kyoto, 606-8501, 
 Japan; 
 email: {\tt nagahara@ieee.org}}%
\thanks{
D. E. Quevedo is with School of Electrical Engineering \& 
 Computer Science, The University of Newcastle, NSW
 2308, Australia;
 email: {\tt dquevedo@ieee.org}}%
\thanks{D. Ne\v{s}i\'{c} is with
 Department of Electrical and Electronic Engineering, 
 The University of Melbourne, Victoria 3010
 Australia; email: {\tt dnesic@unimelb.edu.au}} 
}

\begin{document}
\maketitle
\thispagestyle{empty}
\pagestyle{empty}

\begin{abstract}
In this article,
we introduce a new paradigm of control,
called hands-off control,
which can save energy and reduce CO2 emissions
in control systems.
A hands-off control is defined as a control that has
a much shorter support than the horizon length.
The maximum hands-off control is the minimum support (or sparsest)
control among all admissible controls.
With maximum hands-off control, actuators in the feedback control system can be stopped
during time intervals over which the control values are zero.
We show the maximum hands-off control is given by $L^1$ optimal control,
for which we also show numerical computation formulas.
\end{abstract}

\section{Introduction}
In practical control systems, 
we often need to
minimize the control effort
so as to achieve control objectives
under limitations in equipment such as
actuators, sensors, and networks.
For example, 
the energy (or $L^2$-norm) of a control signal is minimized
to prevent engine overheating
or to reduce transmission cost
with a standard LQ (linear quadratic) control problem;
see e.g., \cite{AndMoo}.
Another example is the \emph{minimum fuel} control,
discussed in e.g., \cite{AthFal},
in which the total expenditure of fuel is minimized with
the $L^1$ norm of the control.

Alternatively, in some situations, the control effort can be dramatically reduced by
holding the control value \emph{exactly zero} over a time interval.
We call such control a \emph{hands-off control}.
A motivation for hands-off control is a stop-start system
in automobiles.
It is a hands-off control; it automatically shuts down 
the engine to avoid it idling for long periods of time.
By this, we can reduce CO or CO2 emissions as well as fuel consumption
\cite{Dun74}.
This strategy is also used in hybrid vehicles \cite{Cha07};
the internal combustion engine is stopped when
the vehicle is at a stop or the speed is lower than a preset threshold,
and the electric motor is alternatively used.
Thus hands-off control is also available for solving environmental problems.
Hands-off control is also
desirable for networked and embedded systems
since the communication channel is not used
during a period of zero-valued control.
This property is advantageous in particular for wireless communications
\cite{JeoJeo06}.
In other words, hands-off control is the least \emph{attention}
in such periods.
From this point of view, hands-off control that maximizes the total time of no attention is somewhat related to the concept of minimum attention control
\cite{Bro97}.

Motivated by these applications, we propose 
a new paradigm of control, called \emph{maximum hands-off control}
that maximizes the time interval over which the control is exactly zero.
Although this type of optimization is highly non-convex,
we have proved in \cite{NagQueNes13} that
under the normality assumption on the optimal control problem,
the maximum hands-off control is given by $L^1$ optimal control,
which can be solved much more easily \cite{AthFal}.

\section{Optimal Control Problems}
\label{sec:problems}
We here consider nonlinear plant models of the form
\begin{equation}
 \frac{d\vc{x}(t)}{dt} = \vc{f}\bigl(\vc{x}(t)\bigr) + \sum_{i=1}^m \vc{g}_i\bigl(\vc{x}(t)\bigr)u_i(t),
 \quad t\in[0,T],
 \label{eq:plant}
\end{equation}
where
$\vc{x}$ is the state,
$u_1,\dots,u_m$ are the control inputs,
$\vc{f}$ and $\vc{g}_i$
are functions on $\R^n$.
We assume that $\vc{f}(\vc{x})$, $\vc{g}_i(\vc{x})$,
and their Jacobians $\vc{f}'(\vc{x})$, $\vc{g}_i'(\vc{x})$
are continuous in $\vc{x}$.
We use the vector representation $\vc{u}\eq[u_1,\dots,u_m]^\top$.

The control $\{\vc{u}(t): t\in[0,T]\}$ is chosen to drive the state $\vc{x}(t)$
from a given initial state 
\begin{equation}
 \vc{x}(0)=\vc{x}_0,
 \label{eq:initial_state}
\end{equation} 
to the origin by a fixed final time $T>0$, that is,
\begin{equation}
 \vc{x}(T)=\vc{0}.
 \label{eq:final_state}
\end{equation}
Also, the control $\vc{u}(t)$ is constrained in magnitude by
\begin{equation}
 \|\vc{u}(t)\|_\infty \leq 1,\quad \forall t \in [0,T].
 \label{eq:input_constraint}
\end{equation}
We call a control $\{\vc{u}(t): t\in[0,T]\}$ \emph{admissible}
if it satisfies \eqref{eq:input_constraint}
and the resultant state $\vc{x}(t)$ from \eqref{eq:plant} satisfies boundary conditions
\eqref{eq:initial_state} and \eqref{eq:final_state}.
We denote by $\U$ the set of all admissible controls.

The \emph{maximum hands-off control}
is a control that
maximizes the time interval over which the control $\vc{u}(t)$ is exactly zero.
In other words, we try to find the \emph{sparsest} control
among all admissible controls in $\U$.

We state the associated optimal control problem as follows:
\begin{problem}[Maximum Hands-Off Control]
\label{prob:MHO}
Find an admissible control $\{\vc{u}(t): t\in[0,T]\}\in\U$ that minimizes
\begin{equation}
 J_0(\vc{u}) \eq \sum_{i=1}^m\lambda_i \|u_i\|_{L^0},
 \label{eq:J_MHO}
\end{equation}
where $\lambda_1>0,\dots,\lambda_m>0$ are given weights.
\end{problem}

On the other hand, if we replace $\|u_i\|_{L^0}$ in \eqref{eq:J_MHO}
with the $L^1$ norm $\|u_i\|_{L^1}$,
we obtain the following \emph{$L^1$-optimal control} problem,
also known as \emph{minimum fuel control}
discussed in e.g. \cite{Ath63,AthFal}.
\begin{problem}[$L^1$-Optimal Control]
\label{prob:L1}
Find an admissible control $\{\vc{u}(t): t\in[0,T]\}\in\U$ that minimizes
\begin{equation}
 J_1(\vc{u}) \eq \sum_{i=1}^m \lambda_i \|u_i\|_{L^1} = \int_0^T \sum_{i=1}^m \lambda_i |u_i(t)| dt,
 \label{eq:J_L1}
\end{equation}
where $\lambda_1>0,\dots,\lambda_m>0$ are given weights.
\end{problem}

\section{Maximum Hands-Off Control and $L^1$-Optimal Control}
\label{sec:main}
In this section, we consider a theoretical relation
between
maximum hands-off control (Problem \ref{prob:MHO})
and $L^1$-optimal control (Problem \ref{prob:L1}).
The theorem below rationalizes the $L^1$ optimality
in computing the maximum hands-off control
\cite{NagQueNes13}.
\begin{thm}
\label{thm:L1optimal}
Assume that the $L^1$-optimal control problem stated in Problem~\ref{prob:L1}
is normal%
\footnote{
When the optimal control is uniquely determined almost everywhere from the minimum principle, the control problem is called normal.
See \cite{AthFal} for details.
}
and has at least one solution.
Let $\U_0^\ast$ and $\U_1^\ast$ be the sets of the optimal solutions
of Problem \ref{prob:MHO} ($L^0$-optimal control problem)
and Problem \ref{prob:L1} ($L^1$-optimal control problem)
respectively.
Then we have $\U_0^\ast = \U_1^\ast$.
\end{thm}

Theorem~\ref{thm:L1optimal} suggests that
$L^1$ optimization can be used for 
the maximum hands-off (or the sparsest) solution.
This is analogous to the situation in compressed sensing,
where $L^1$ optimality is often used to obtain the sparsest vector;
see \cite{HayNagTan13} for details.

\section{Linear Plants and Numerical Computation}
\label{sec:computation}
We here propose a numerical computation method
to obtain an $L^1$-optimal control (i.e. maximum hands-off control)
when the plant model is linear and time-invariant.

Let us consider the following linear time-invariant plant model
\begin{equation}
 \frac{d\vc{x}(t)}{dt} = A\vc{x}(t) + B\vc{u}(t),\quad t\in[0,T], \quad \vc{x}(0)=\vc{x}_0,
 \label{eq:plant_ln}
\end{equation}
where $\vc{x}(t)\in\R^n$ and $\vc{u}(t)\in\R^m$.
We assume that the initial state $\vc{x}_0\in\R^n$ and
the time $T>0$ are given.

Linear systems are much easier to treat than general nonlinear systems
as in \eqref{eq:plant}.
In particular, 
for special plants, such as single or double integrators,
the $L^1$-optimal control can be obtained analytically;
see e.g., \cite[Chap.~8]{AthFal}.
However, for general linear time-invariant plants,
one should rely on numerical computation.
For this, we adopt a time discretization approach
to solve the $L^1$-optimal control problem.
This approach is standard for numerical optimization;
see e.g.~\cite[Sec.~2.3]{Ste}.

We first divide the interval $[0,T]$ into $N$ subintervals,
$[0,T] = [0,h) \cup \dots \cup [(N-1)h,Nh]$,
where $h$ is the discretization step chosen such that $T=Nh$.
We here assume (or approximate) that the state $\vc{x}(t)$ and the control $\vc{u}(t)$ are
constant over each subinterval.
On the discretization grid,
$t=0,h,\dots,Nh$,
the continuous-time plant \eqref{eq:plant_ln} is described as
\[
 \vc{x}_d[m+1] = A_d\vc{x}_d[m] + B_d \vc{u}_d[m],\quad m=0,1,\dots,N-1,
\]
where $\vc{x}_d[m]\eq\vc{x}(mh)$, $\vc{u}_d[m]\eq \vc{u}(mh)$, and
\[
 A_d \eq e^{Ah},\quad B_d \eq\ \int_0^h e^{At}Bdt.
\]
Set the control vector
\[
 \vc{U}\eq[\vc{u}_d[0]^\top,\vc{u}_d[1]^\top,\dots,\vc{u}_d[N-1]^\top]^\top.
\] 
Note that the final state $\vc{x}(T)$ can be described as
\[
 \vc{x}(T)=\vc{x}_d[N]=A_d^N \vc{x}_0 + \Phi_N\vc{U},
\]
where
\[
 \Phi_N \eq \begin{bmatrix}A_d^{N-1}B_d,&A_d^{N-2}B_d,&\dots,&B_d\end{bmatrix}.
\]
If we define the following matrices:
\[
 \begin{split}
  \Lambda_m &\eq \diag(\lambda_1,\dots,\lambda_m),~
  \Lambda \eq \blockdiag(\underbrace{\Lambda_m,\dots,\Lambda_m}_{N}),
 \end{split}
\]
then the $L^1$-optimal control problem
is approximately described as
\begin{equation}
 \begin{aligned}
  & \underset{\vc{U}\in\R^{mN}}{\text{minimize}}
  & & \|\Lambda\vc{U}\|_1\\
  & \text{subject to}
  & & \|\vc{U}\|_\infty\leq 1,~
  A_d^N\vc{x}_0 + \Phi_N\vc{U}=\vc{0}.
 \end{aligned}
 \label{eq:l1optimization}
\end{equation}
The optimization problem \eqref{eq:l1optimization}
is convex and can be efficiently solved by
numerical software packages such as \verb=cvx= with MATLAB;
see \cite{cvx} for details.

\section{Conclusion}
\label{sec:conclusion}
In this article, we have 
presented maximum hands-off control and shown that it is $L^1$ optimal.
This shows that efficient optimization methods for $L^1$ problems can be used to obtain maximum hands-off control.
A time discretization method has been presented for the computation of
$L^1$-optimal control when the plant is linear time-invariant.
The resultant optimization is a convex one,
and hence can efficiently be solved.
Future work may include adaptation of hands-off control to sparsely packetized predictive control
as in \cite{NagQue11,NagQueOst13}.



\end{document}